\documentclass[useAMS,usenatbib]{mn2e}

\usepackage{graphicx}	

\title[New TiO Index]{A New Titanium Oxide Index  in Visual Band}

\author[B. Bidaran et. al.]{Bahar Bidaran$^{1}$\thanks{E-mail: b.bidaran@student.alzahra.ac.ir (BB)}, Mohammad Taghi Mirtorabi$^{1}$ and Fatemeh Azizi$^{2} $\\
$^{1}$Department of Physics, Alzahra University,  P.O. Box 1993893973, Tehran, Iran\\
$^{2}$Department of Physics, Payame Noor University (PNU), P.O. Box 19395-3697, Tehran, Iran}

\begin{document}
\label{firstpage}
\pagerange{\pageref{firstpage}--\pageref{lastpage}}
\maketitle

\begin{abstract}
We introduce a new color index consisting of two spectral band width to measure TiO absorption band strength centered at 567 nm. Based on the most up-to-date line list for TiO we regenerate a grid of synthesized  spectra and investigate the temperature sensitivity of the index. The new index behave similar to older Wing TiO-index where it decrease monotonically from coolest atmosphere with $T_{eff} = 2800$ up to $T_{eff} = 4000$ where the TiO molecules disassociate. To further examine the feasibility of the new index we reproduce the calibration using a list of observed high resolution spectra and found similar results. This index extend TiO absorption band capability to measure effective temperature of late K to M stars to visual spectrum where it is more accessible to small telescopes for long term dedicated observation.
\end{abstract}

\begin{keywords}
line formation - molecular data - late-type stars - instrumentation:spectrograph.
\end{keywords}



\section{Introduction}

Molecular absorption is a signature of cool atmosphere in late type stars and magnetized active regions in F - K main sequences. They are consisting of wide band absorptions, observable with low resolution spectrographs or simple photometers attached to small telescopes. In a cool star atmosphere, atomic lines are weaker because of atomic depletion by molecules or grains formation or lower atomic states population. Due to physical conditions such as low temperature and low density which can cause infrequent collisions, the local density of molecules such as ZrO, CO, TiO and VO grow in cool atmospheres, because they survive from thermal dissociation \citep{Rob07}. Molecular absorption may influence the spectrum as an individual lines or suppress the continuum in a wide interval of wavelengths which can change the whole structure of the spectrum. These pervasive spectral features have been used as a major characteristic of cool atmospheres to classify late type M stars \citep{Ramsey}.

Among molecules mentioned above, TiO comes across as special one due to its wide and strong absorption in near infra-red. Its absorption bands start appearing and deepening in late K and prevail until coolest M type stars. In order to study cool stars using TiO, Wing has introduced a photometric system consists of three filters which are tuned to measure TiO $\gamma(1,1)$ R-branch band head at 719 nm on near infra-red \citep{Wing}. He proposed this region because it was accessible to the solid state photometers at the time. This photometric system was used to study the correlation between choromospheric activity and TiO absorption strength in variable star $\lambda$ Andromeda \citep{Mirtorabi}. They have found an estimated activity cycle of about 4 to 14 years and an anti-correlation between average TiO absorption representing coverage of spots with average visual brightness; similar to what is observed in the Sun. The most updated continuum band-passes of this system has been published recently by \citet{Azizi}. They have used observed high resolution spectra to investigate whether Wing continuum band passes are influenced with other molecular absorption then updated the photometric system. In the updated system the continuum band passes are shifted to shorter wavelengths where the correlation between absorption strength and effective temperature shows less scatter. Based on Wings system, Ramsey also calibrated TiO$\lambda$886 nm absorption band as a function of effective temperature \citep{Ramsey}.

\begin{figure*}
\includegraphics [width=15cm]{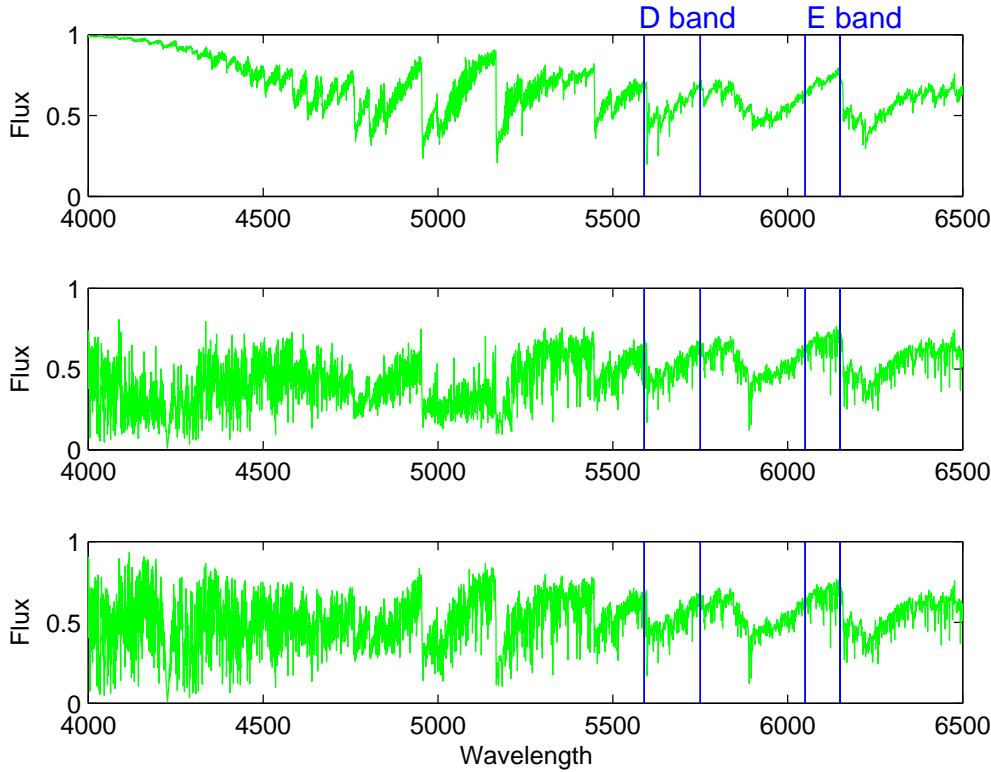}
\caption{Synthetic Spectra of a star with $T_{eff}$ = $3750 K$ and $\log g$ = $3.0$ normalized to continume. The top panel shows absorption features belongs to  TiO molecules with all other molecules and atoms omitted. In the middle panel contaminations produced by other molecules to TiO absorption bands are added. In the bottom panel all the molecules and atoms absorption lines are included. The selected band passes D and E are also shown on the panels.}
\label{fig:01}
\end{figure*}

During 90's on one hand, extended TiO line lists were released which overcome the computational limitations especially in the visual region, on the other hand, large spectroscopic surveys made a vast amount of observed spectra of stars accessible to astrophysicist. These improvements have made it possible to search for TiO absorption bands in the  visual region. We were motivated by \citealp{Mayor}, where they have searched for star's activity identification based on Ca II absorption using HARPS spectrograph at ESO. We have selected a list of K to  M stars observed with  KPNO 2.1 m telescope to investigate detectable TiO absorption in visual region. We used these spectra to make a calibration between TiO absorption and effective temperature. In section 2 we describe basic properties of TiO lines in visual region. In section 3 we introduce our index and later on section 4 we calibrated this index with effective temperature of K-M stars using their observed spectra.

\section{TiO absorption in visual wavelengths}
Although as a heavy metal, Titanium has tiny abundance in atmosphere of stars but Titanium Oxide is the most pervasive spectroscopic feature of cool stellar atmospheres in near infra-red region. Titanium Oxide molecule form in temperatures lower than 4000 K. TiO opacity increases by decreasing temperature down to coolest atmosphere ever known then it can be used as a surface temperature indicator for spectral classes later than K \citep{Wing}. TiO transitions between its ground and upper excited states in optical and near infrared wavelengths  occurs mainly between electronic systems and can be divided into two groups: the first group are called allowed transitions and the strongest ones are $\gamma$($A^3 \Delta-X^3\Delta$), $\gamma'$($B^3 \Pi-X^3 \Delta$) and $\alpha$($C^3 \Delta-X^3 \Delta$), in which all the A,B and C corresponds to excited levels with the same multiplicity as its ground state. The second group which can be referred as forbidden transitions are those occur in transitions between ground state and the non-equal multiplicity excited state. The strongest transitions of this group are: $\beta$($c^1 \Phi-a^1 \Delta$), $\delta$($b^1 \Pi-a^1 \Delta$)and $\phi$($b^1 \Pi-d^1 \Sigma^{+}$) \citep{Dobrodey}. The first group transitions are normally stronger than the second. Other transitions such as vibrational, rotational and transitions due to non-zero spin may influence TiO's spectrum marginally.

\begin{figure*}
\includegraphics [width=15cm]{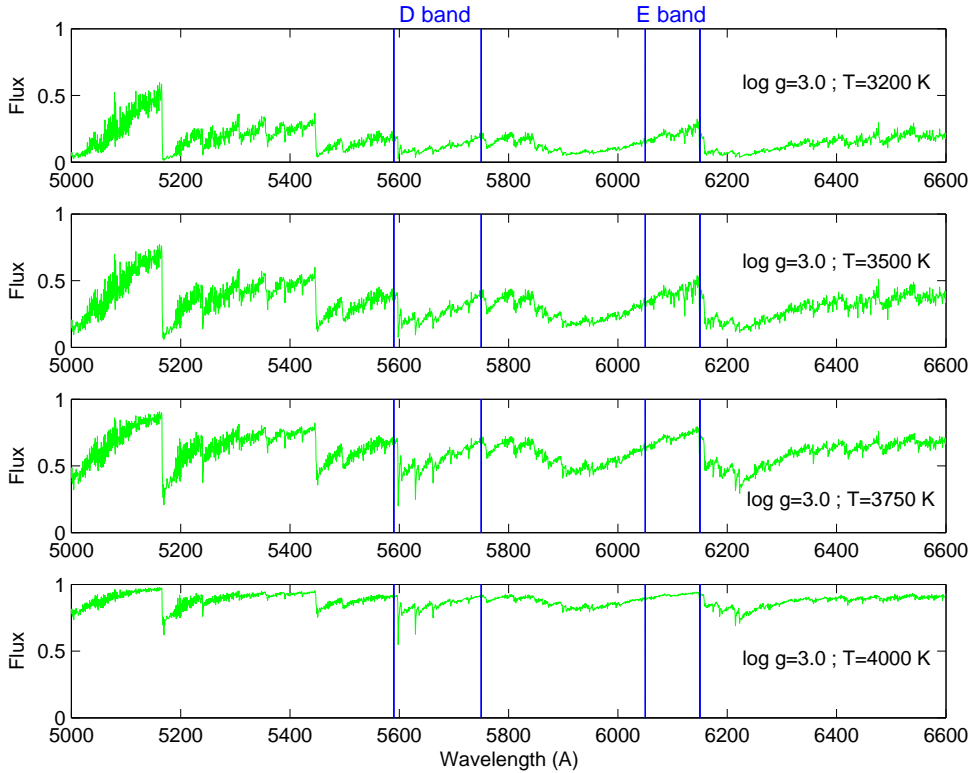}
\caption{TiO absorption bands in four different temperatures with the same gravity. All the data has been produced using ATLAS 9 software. As it is obvious from the figure, TiO absorption bands deepens as temperature decreases declares that TiO's population increases as the star becomes cooler.The selected band passes  D and E are also shown in the panels. In all four panels flux is normalized to continume.
Filter D measures 559-575(nm).filter E measures 605-615(nm).}
\label{fig:02}
\end{figure*}

 line lists calculated for TiO we found Schwenke's the most precise and suitable one for atmospheric modeling. He has computed a combined list completed to high energies \citep{Allard}. This line lists includes 172 million lines. About 32 millions of them corresponds to four isotopes of TiO ($^{50,49,47,46}$TiO), and 48 millions  belongs to the most abundant one ($^{48}$TiO) \citep{Allard}. Figure \ref{fig:01} shows three synthetic spectra of a star with $T_{eff} = 3750 K$ and $\log g = 3.0$ calculated using ATLAS9 to show the effect of molecular and atomic absorption on the spectrum in visual wavelengths. We assumed local thermodynamic equilibrium (LTE) for the whole line forming region. \citet{Valenti1998} has checked that this is a good assumption for the great portion of line formation region although the absorption band core form in the high atmosphere where LTE is weak. To find a wide absorption  feature capable of reproducing observable spectral index, we calculated a spectrum with all atomic and molecular absorption excluded except TiO.The top panel of figure \ref{fig:01} shows absorption features of the sole TiO molecules in visual region. In the middle panel we add other molecular absorptions to the spectrum to explore possible contamination of TiO absorption systems. It is clear that TiO absorption bands in wavelengths shorter 5500 nm are strongly affected by presence of other molecules. As the middle panel shows the  absorption system centered on 567 nm is less influenced  and it can represent TiO absorption even in presence of other molecules. To be sure that atomic absorption will not destroy this system either,  we finally turn on atomic absorption lines which is shown in the bottom panel. Figure \ref{fig:01} reveals that except the one at 567 nm, other shorter absorption features of TiO are completely washed out by  molecules like H-O, CO and CN.  \\
 The effect of vanadium oxide molecules in visual part of spectrum is negligible. \cite{Kirkpatrick} has listed absorption bands of this molecules with the shortest  system at 733.4 nm, which is far from our selected system. Finally we can conclude that the TiO absorption system  at 567 nm is less contaminated by the molecular and atomic bands and could be used as a TiO absorption representative. This band is a $\beta$ (2-2) system of TiO 's 48 isotope consisting of 48 lines \citep{Valenti1998}. 

\section{Visual TiO index}

A feasible spectral index which is going to represent abundance of a molecular junction or capable of calibrating a physical parameter   like the  effective temperature basically compares two region of the spectrum one deep inside the absorption region and the other far out on the continuum. The absorption feature might be wide enough to be observable with low resolution spectrographs or simple filter based photometers to admit long term dedicated observation with small telescopes. To explore the visual region for such TiO absorption we calculated atmospheric models for a grid of cool giant and dwarf stars using ATLAS 9 software package \citep{Kurucz1991}. The grid consists of atmospheric models ranging from $2900 \leq T_{eff} \leq 4200 K$ with steps $\Delta T = 100 K$. The surface gravity was varies in the interval $3 \leq \log g \leq 5$ with steps $\Delta g = 0.5$ \citep{Castelli1997}. A number of 70 synthetic spectra were obtained.
Figure \ref{fig:02} shows four spectra of this sample corresponds to $\log g = 3.0$ with different effective  temperatures.
To search for a continuum band we can find  at best only three regions to be labeled as continuum in visual wavelength in Figure \ref{fig:02}: The first part (630-660) nm  is too contaminated to be labeled as continuum. The second part (520-540) nm seems to become noisy considering other molecules and atomic  effect( Figure \ref{fig:01}). As we need the continuum not to be crowded with other overlapping TiO absorption bands, we cannot use this region due to some $\beta$ systems of TiO covers this part of  spectrum \citep{Valenti1998}. It is also important to have a well-defined continuum near our absorption system to make the measurement of band head more precise.The third availabe continuum which is in(605-615) nm  seems to be the most relevant one close to our porposed TiO band and free of other molecular absorption.

\begin{figure}
\includegraphics[width=\columnwidth]{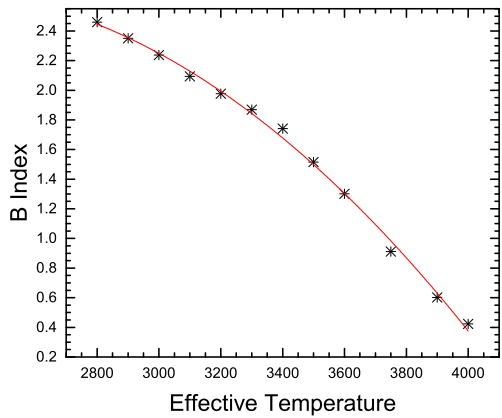}
\caption{TiO B-index vs. effective temperature for synthetic spectra. all the data have been produced with ATLAS9.}
\label{fig:03}
\end{figure}
 
\begin{figure}
\includegraphics[width=\columnwidth]{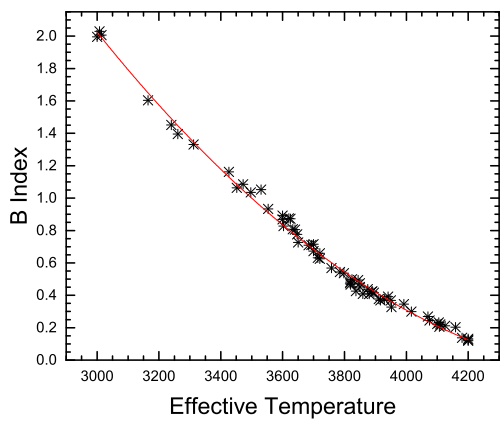}
\caption{ TiO B-index vs. effective temperature for observed spectra. All the information obtained from KPNO.It is clear that for real stars we observe a similar trend for the B-index as we have for synthesized spectra.}
\label{fig:04}
\end{figure}

Following Wing's system, we introduced two new filters D and E. The central wavelengths and full width at half-maximum band pass at each filters are listed in Table~\ref{Tab:01}. Filter D measure the TiO absorption band strength centered at 567 nm. The width of filter D is set to 16 nm which is widest possible to make use of whole absorption band.  Filter E is centered at 610 nm with a width of 10 nm.  A new visual Titanuim index  which we call it "B-INDEX" can be formed by comparing integrated flux observed from these two band passes. By the way we define $B$ - index as 
  
\begin{equation}
B-index = - 2.5 \log \frac{\int F_D(\lambda) S_D(\lambda)d\lambda}{\int F_E(\lambda) S_E(\lambda)d\lambda}  
\label{eq:03}
\end{equation}

where $F_D(\lambda)$ and $F_E(\lambda)$ are integrated spectral flux in each band passes and $S_D(\lambda)$ and $S_E(\lambda)$ are appropriate filter response functions in each filter. 

To demonstrate temperature sensitivity of the B-index, we assume a flat response function for both filters and calculate integrated flux in each filter for all synthesized spectra.  Although the absorption band in 567 nm has a small signature  of atomic absorption or possible molecular contamination, Figure~\ref{fig:03} shows that the trend of variation of the index with  effective temperature is smooth and monotonic.

\begin{table}
\begin{tabular}{@{}lcccc}
\hline
Filter & Region Measured & Central & Bandpass\\
& & wavelength & FWHM\\
& & (nm) &(nm)\\
\hline
D &TiO$\lambda$567 nm band & 567 & 16\\
E & continum & 610 & 10\\
\hline
\end{tabular}
\caption{ The B-index filter's system. }
\label{Tab:01}
\end{table}

\begin{table*}
\begin{tabular}{@{}lcccccccccc}
\hline
HD & TiO-Index (mag) & $T_{eff} $ (K)& Sp. Type& Ref. & HD & TiO-Index  (mag) & $T_{eff}$ (K) & Sp. Type& Ref. \\
\hline
147379 & 0.3763 & 3929 & M1 & [3] & 20797 & 0.5686 & 3758 & M0 &[2]\\
29139 & 0.4520 & 3850 & K5 & [2] & 34255 & 0.3776 & 3911 & K4 & [5]\\
34334 & 0.1370 & 4180 & K2 & [6] & 35620 & 0.1309 & 4200 & K3 & [10]\\
39225 & 0.6257 & 3720 & M2 & [7] & 39801 & 0.8077 & 3633 & M1-M2 & [11]\\
39853 & 0.4253 & 3837 & K5 & [10] & 40239 & 1.0526 & 3530 & M3 & [7]\\
42543 & 0.7095 & 3684 & M0 & [2] & 44478 & 0.8693 & 3600 & M3 & [9]\\
44537 & 0.5364 & 3798 & K5 & [1],[7] & 57651 & 0.6321 & 3714 & K5 & [10]\\
78712 & 1.6038 & 3165 & M6 & [10] , [11] & 80493 & 0.4082 & 3880 & K7 & [10]\\
83787 & 0.4743 & 3818 & K6 & [2] & 97907 & 0.1237 & 4200 & K3 & [12]\\
99167 & 0.4813 & 3819 & K5 & [2] & 99998 & 0.2121 & 4125 & K3.5 & [1]\\
102212 & 0.4946 & 3844 & M1 & [16] & 108985 & 0.3691 & 3950 & K5 & [7]\\
111355 & 0.438 & 3875 & K5 & [2] & 112142 & 0.7764 & 3647 & M3 & [9]\\
112300 & 0.8754 & 3620 & M3 & [13] & 114961 & 2.0070 & 3014 & M7 & [10]\\
114287 & 0.3459 & 3992 & K5 & [2] & 118100 & 0.2230 & 4100 & K5 & [14]\\
120933 & 0.8721 & 3625 & K5 & [1] & 123657 & 1.0631 & 3452 & M4.5 & [15]\\
126327 & 1.9960 & 3000 & M7.5 & [15] & 123934 & 0.544 & 3785 & M1 & [1] ,[2]\\
128000 & 0.3918 & 3941 & K5 & [17] & 131918 & 0.2056 & 4107 & K4 & [5],[2],[16],[17],[18],[4]\\
136726 & 0.2034 & 4159 & K4 & [20] & 138481 & 0.4247 & 3890 & K5 & [19]\\
139669 & 0.4161 & 3895 & K5 & [20] & 175588 & 1.0856 & 3472 & M4 & [1]\\
148513 & 0.2447 & 4075 & K4 & [10] & 148783 & 1.3954 & 3261 & M6 & [1]\\
149161 & 0.3277 & 3952 & K4 & [16] & 158899 & 0.22 & 4109 & K4 & [22]\\
167006 & 0.8054 & 3640 & M3 & [4] & 168720 & 0.4663 & 3820 & M1 & [1]\\
180928 & 0.2322 & 4105 & k4 & [1],[2],[5] &169305 & 0.7149 & 3700 & M2 & [1],[7]\\
177940 & 2.0292 & 3008 & M7 & [1] & 186619 & 0.4220 & 3888 & M0 & [1],[2]\\
189319 & 0.3683 & 3916 & M0 & [10] & 191372 & 0.7264 & 3650 & M3 & [7]\\
196610 & 1.4513 & 3240 & M6& [7] & 197812 & 1.3308 & 3312 & M5 & [1]\\
196777 & 0.6598 & 3720 & M1 & [7] & 197939 & 0.8274 & 3603 & M3 & [5]\\
200527 & 0.9327 & 3553 & M4 & [1] & 202259 & 0.4990 & 3824 & M1 & [1],[5]\\
203535 & 0.4093 & 3859 & M0 & [1] & 204445 & 0.7110 & 3693 & M1 & [1],[7]\\
206936 & 0.6733 & 3700 & M2 & [19] & 207134 & 0.1195 & 4200 & K3 & [12]\\
217906 & 0.8917 & 3600 & M2 & [10] & 221615 & 1.1615 & 3426 & M5 & [1]\\
225212 & 0.2988 & 4016 & K3 & [22],[28] & 237903 & 0.2694 & 4070 & K7 & [17]\\
146051 & 0.4751 & 3850 & M0.5 & [21] & 172816 & 1.0358 & 3497 & M4 & [10]\\
\hline
\end{tabular}
\caption {List of stars. References: [1] \citet{Rimoldini}; [2] \citet{Gontcharov}; [3] \citet{ShayaO}; [4] \citet{McDonald}; [5]\ \citet {Bobylev}; [6] \citet{Ofek}; [7] \citet{Teyssier}; [8] \citet{Ramirez}; [9]\citet{Smith}; [10] \citet{Blum}; [11] \citet{Glebocki}; [12]\citet{Borgne}; [13] \citet{Salim}; [14] \citet{Cenarro} ; [15] \citet{Price} ; [16] \citet{Koleva} ; [17] \citet{Anderson} ; [18] \citet{Sanchez} ;[19] \citet{Eggleton} ; [20] \citet{Massarotti} ; [21] \citet{Tabur} ; [22] \citet{Wallace}. }
\label{Tab:02}
\end{table*}

To evaluate this index with observed spectra  we selected a list of 78 stars with spectral index spanning from K2 to M8. These stars were observed with the Kitt Peak National Observatory (KPNO) 2.1 m telescope. The telescope was occupied with coud\'{e} Feed spectrograph with a F3KB CCD with pixel size of 15 micron which span visual wavelengths range from 346.5 to 946.9 nm. They are listed in the first and fifth column of Table~\ref{Tab:02}. Among these stars 12 was discarded due high level of noise or incomplete data. Effective temprature for these stars were found by searching on available observation  or related literatures which is mentioned in fifth and tenth coloumn of  Table~\ref{Tab:02}. By the way the new defined B - index was calculated for all of these spectra to form a B-index-temperature calibration. Figure \ref{fig:04}  shows that similar to synthesized spectra a smooth and monotonic decrease is seen which clearly shows that the  new defined index has capability to recognize the effective temprature  of cool atmospheres. To illustrate the calibration relation we fit a  second order polynomial to the plot and found a simple calibration relation 

\begin{equation}
B-index = 6.667\times 10^{-7} T_{eff}^{2} - 0.0064 \times T_{eff} + 15.17\label{eq:02}
\end{equation}

The standard deviation of the residuals around the fit is 0.0719. As it can be seen from the curve, the value of B-index is declining with increasing temperature. By decreasing temperature the TiO index increases up to a value about 2.00 which is corresponded to spectral type M7.5.\\

\section{Discussion}
Figure \ref{fig:04} shows that the new B-index mimics the original TiO-index invented by Wing. The reader may ask why we need a visual representative for cool atmosphere of K-M stars where they are intrinsically faint in their visual spectrum. The original TiO-index was adopted to the one of the strongest TiO absorption bands in the near infra-red where the spectrum is  bright in $2900 \leq T_{eff} \leq 4200 K$. \cite{Mirtorabi}  have shown that TiO-index can reveal stellar activity of F-G type stars. They observed $\lambda$ And in near infra-red and detected TiO absorption through nonzero and variable Wing TiO-index. The quiet photosphere of $\lambda$ And is too hot to have TiO molecule. It is believed that part of the variability of this star comes from large coverage of unevenly distributed spots.  The absorption might come from this cool active regions. They also found that TiO molecules can survive even when the star is brightest which means part of the active regions might distributed evenly on the surface of the star with no observable effects on visual light curve. A visual TiO index can represent this activity more precisely. In future work we are going to use the B-index to search for stellar activity in visual spectrum of F and G stars where they are more brighter in visual than infra-red.

\label{lastpage}
\end{document}